\documentclass[aps,pra,reprint,amsmath,amssymb,superscriptaddress,floatfix,showpacs,showkeys,subeqn]{revtex4-1} 
\usepackage{hyperref}
\usepackage{graphics} 
\usepackage{graphicx} 
\usepackage{subfigure}
\usepackage[normalem]{ulem}
\usepackage{url}
\usepackage[utf8]{inputenc}
\newcommand{\rme}{\mathrm{e}}
\newcommand{\rmi}{\mathrm{i}}

\newcommand{\field}[1]{\mathbb{#1}} 

\begin{document} 
\author{L. A. Toikka}
\affiliation{Dodd-Walls Centre for Photonic and Quantum Technologies, Centre for Theoretical Chemistry and Physics, and New Zealand Institute for
Advanced Study, Massey University, Private Bag 102904 NSMC, Auckland 0745, New
Zealand}
\author{K.-A. Suominen}
\affiliation{Turku Centre for Quantum Physics, Laboratory of Quantum Optics, Department of Physics and Astronomy, University of Turku, FI-20014 Turku, Finland}
\title{Analytic Vortex Dynamics in an Annular Bose-Einstein Condensate}
\date{\today}
\begin{abstract}
We consider analytically the dynamics of an arbitrary number and configuration of vortices in an annular Bose-Einstein condensate obtaining expressions for the free energy and vortex precession rates to logarithmic accuracy. We also obtain lower bounds for the lifetime of a single vortex in the annulus. Our results enable a closed-form analytic treatment of vortex-vortex interactions in the annulus that is exact in the incompressible limit. The incompressible hydrodynamics that is developed here paves the way for more general analytical treatments of vortex dynamics in non-simply connected geometries.
\end{abstract}
\keywords{vortex dynamics, free energy, Bose-Einstein condensate, precession, ring trap}
\maketitle

\section{Introduction}
The experimental observation of quantised persistent currents in toroidally confined Bose-Einstein condensates (BECs)~\cite{PhysRevLett.99.260401, PhysRevLett.106.130401, PhysRevA.86.013629} is closely related to the appearance of vortices with quantised circulation~\cite{RevModPhys.81.647, PhysRevLett.85.2857}. The dynamics of quantised vortices and dark solitons~\cite{Greekreview,PhysRevLett.84.2298,PhysRevLett.86.2926} in BECs is a key ingredient in the understanding of the properties of superconductors~\cite{tilley1990superfluidity} and quantum liquids~\cite{Donnelly91}, and have been intensely studied by the BEC community~\cite{RevModPhys.73.307, leggett2006quantum}. 

The persistent current state is protected by an energy gap, preventing the creation of excitations at low energies. Above the Landau critical velocity, which equals to the speed of sound in a homogeneous condensate, the superfluid state becomes unstable against the formation of excitations such as vortices~\cite{PhysRevLett.82.5186} and dark solitons~\cite{PhysRevLett.99.160405} in the bulk. The behaviour of these collective non-linear excitations can be distinctively different in non-simply connected geometries~\cite{PhysRevA.79.043620, ToikkaSuominen2014} as a result of the `holed' topology. For example, in the ring geometry, the outer and inner surfaces play important separate roles in the vortex nucleation above the critical rotation~\cite{PhysRevA.86.011604, PhysRevA.86.011602}. The ring geometries are within experimental reach~\cite{PhysRevA.74.023617,Griffin2008,PhysRevA.88.063633}, for example, in a genuinely non-trivial ring topology where the central region is occupied by a tapered optical fibre and therefore essentially at infinite potential~\cite{1367-2630-10-11-113008}. The ring traps can be highly flexible with a tunable radius and transverse frequency~\cite{1367-2630-10-4-043012}, making them ideal for the study of multiply-connected BECs. 

In a ring trap sustaining a persistent current, a tunable weak link across one side of the torus has been used to shed vortices~\cite{PhysRevLett.106.130401, PhysRevA.91.023607,2015arXiv1512.07924}. In another experiment, an annular condensate was stirred with a narrow blue-detuned optical beam resulting in vortices appearing in the bulk~\cite{PhysRevA.88.063633}. The weak link alters the local critical velocity affecting the current in the ring, and forms a closed-loop atom circuit with applications in atomtronics~\cite{PhysRevLett.111.205301}.  In annular geometry, many-vortex states such as a circular array of vortex-antivortex pairs can also be produced via the snake instability of a ring dark soliton~\cite{PhysRevA.87.043601,0953-4075-47-2-021002,Toikka2013}, whose subsequent vortex dynamics has been studied in a circular container~\cite{PhysRevA.70.043624,PhysRev.161.189}, and in a harmonic plus quartic trap~\cite{PhysRevA.72.023619,PhysRevA.73.013614}. It was found that when the BEC is rotated sufficiently rapidly, an annular geometry with a vortex lattice that evolves into a ring of vortices can emerge~\cite{PhysRevA.73.013614}.

In a harmonic trap, an off-axis vortex precesses in the same sense as the vortex circulation, following a circular trajectory~\cite{RevModPhys.81.647}. This effect can be explained in terms of a Magnus force~\cite{PhysRevB.55.485,PhysRevA.62.063617,PhysRevA.61.013604} that is pointing radially inwards. The Magnus effect is a transverse force that always appears when a vortex moves with respect to a superfluid~\cite{PhysRevLett.76.3758}, which has been linked to the concept of a geometric Berry phase~\cite{PhysRevLett.70.2158}. In an infinite homogeneous neutral superfluid at zero temperature (in the absence of any quasi-particle scattering~\cite{PhysRevLett.81.4276}), the transverse Magnus force relies on the superfluidity and single-valuedness of the condensate order parameter~\cite{PhysRevLett.70.2158}. Vortex precession is a key observable feature of vortex dynamics, and here we seek to provide closed-form analytical results for vortex dynamics in the ring geometry, including precession in a ring trap.

Another approach for studying vortex dynamics is formed by the method of images~\cite{Donnelly91,PhysRevA.77.032107,PhysRevA.74.043611}, where we assume that the spatial vortex separation is much larger than the healing length. Replacing the condensate boundaries with image vortices provides a direct method for describing the solution to the hydrodynamic boundary value problem in the incompressible Thomas-Fermi limit. The method of images has been studied numerically for an annulus~\cite{PhysRevA.66.033602,PhysRevA.64.063602,PhysRevA.79.043620}; here, we develop closed-form analytical expressions for key observable features describing an arbitrary number and configuration of vortices in a ring BEC that solve the boundary value problem exactly. We obtain expressions for the free energy and vortex precession rates to logarithmic accuracy. Our energy agrees with that obtained by Fetter~\cite{PhysRev.153.285} in studying the critical rotation for exciting vortices in an annulus. On the other hand, our expression for the vortex precession rate, an exact solution in the Thomas-Fermi limit, is qualitatively similar to previously numerically and analytically approximately obtained precession rates~\cite{PhysRevA.66.033602}. As an application, these results are used to obtain a lower bound for the lifetime of a single vortex in a ring-shaped condensate. Furthermore, we show how the Uniqueness Theorem implies that the Magnus force approach gives equivalent results with the method of images for a non-dissipative system in a flat-bottom box trap.

The aim of this work is to elucidate observable aspects of vortex dynamics in an annulus in terms of closed-form results valid for an incompressible two-dimensional superfluid. Starting from the incompressible hydrodynamics presented here, it is possible to generalise the results to spin-orbit coupled and compressible superfluids in non-simply connected geometries. The presentation is organised as follows: In Section~\ref{sec:2a}, we derive an expression for the free energy of an arbitrary number and configuration of vortices in an annulus. In Section~\ref{sec:2b}, we apply the results to derive a closed-form expression for the precession rate of a vortex in an annulus. Unlike in a cylindrical container, the vortex can precess in either clockwise or counter-clockwise sense depending on its radial displacement. In Section~\ref{sec:2c}, we derive lower bounds for the lifetime of a single vortex in a ring trap. We defer technical details to the Appendices.

\section{\label{sec:2}Vortex in an annulus}
\subsection{\label{sec:2a}Free energy} We wish to obtain the free energy $E$ of having $N_v$ vortices at $\lbrace \boldsymbol{\rho}_j \rbrace\, (j = 1,2,\ldots N_v)$ in the annulus $\Omega = \{a < r < b; 0 \leq \varphi < 2\pi\}$. Throughout this work we scale our units so that $\hbar = 2m = 1$, which means we measure time, length, and energy in terms of $\omega_x^{-1}$, $a_{\mathrm{osc}} = \sqrt{\hbar/2m\omega_x}$, and $\hbar \omega_x$ respectively, where $\omega_x$ is the angular frequency of the trap in the $x$-direction, and $m$ is the mass of the atoms in the cloud. First focussing on a single vortex at $\boldsymbol{\rho} \in \Omega$, we will work in terms of a stream function $\chi_{\boldsymbol{\rho}}(\textbf{r})$ such that the superfluid velocity $\textbf{v}$ is given by
\begin{equation}
\label{eqn:sf_vel}
\textbf{v}(\textbf{r}) = \frac{\hbar}{m}\hat{\textbf{z}} \times \nabla \chi_{\boldsymbol{\rho}}(\textbf{r}).
\end{equation}
It follows by definition of the vorticity $\nabla \times \textbf{v}$ of irrotational flow that the stream function satisfies the Poisson equation. To take into account the physical no-flow requirement $n_0 \textbf{v}\cdot \hat{\textbf{n}} = 0$ and the geometry of the system, where $n_0$ is the superfluid density and $\hat{\textbf{n}}$ the unit outward normal, we require Dirichlet boundary conditions: 
\begin{equation}
\label{eqn:Poisson_stream}
\begin{cases}
\nabla^2 \chi_{\lbrace \boldsymbol{\rho}_j \rbrace}(\textbf{r}) = 2\pi \sum_{j=1}^{N_v} \delta(\textbf{r}-\boldsymbol{\rho}_j), \\
\chi_{\lbrace \boldsymbol{\rho}_j \rbrace}(a) = \chi_{\lbrace \boldsymbol{\rho}_j \rbrace}(b) = 0.
\end{cases}
\end{equation}

\begin{figure}
  \centering
  \includegraphics[width=0.45\textwidth]{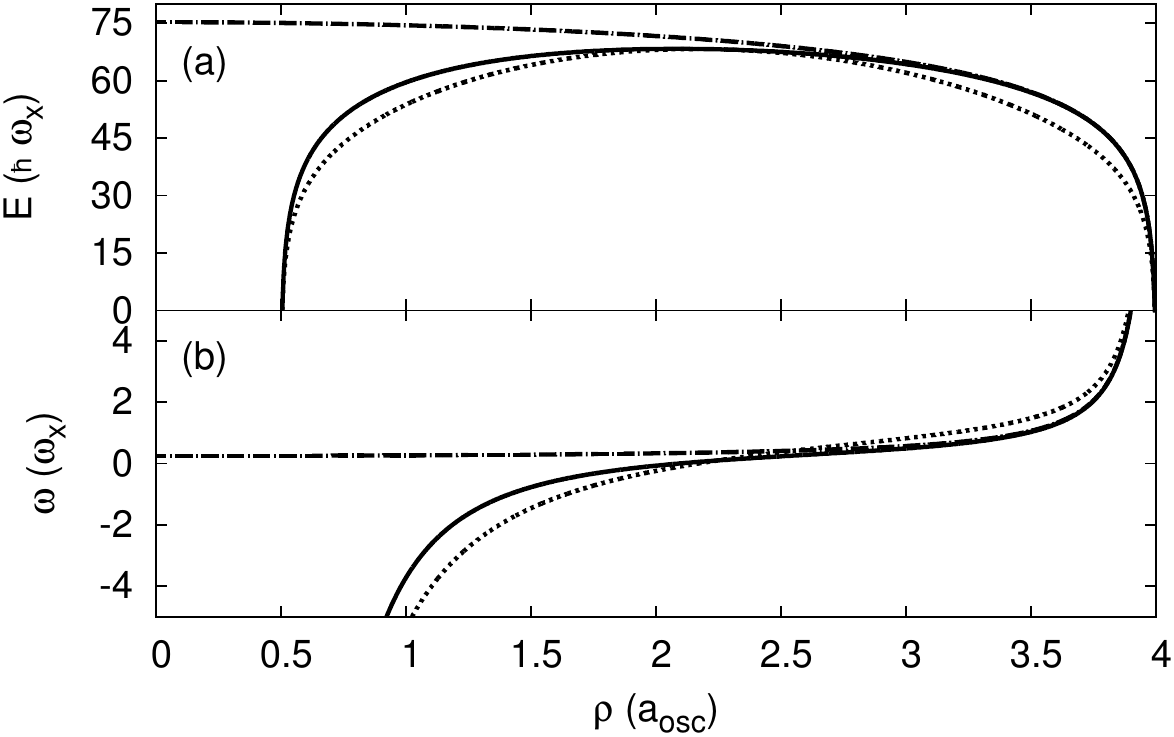}
  \caption{(a) The free energy of a vortex in a uniform annular condensate, Eq.~\eqref{eqn:2en}, with $n_0 = 1$, $\xi = 0.01$, $a = 0.5$, and $b = 4.0$ (solid line). Also the Thomas-Fermi result is shown (dotted line) with $r_{\mathrm{trap}} = (a+b)/2$. (b) The corresponding precession rates around the annulus, Eqs.~\eqref{eqn:uniform_annulus} and~\eqref{eqn:TF_annulus}. There is a quasi-stable radius at which $\omega = 0$ corresponding to the stationary point of the energy. The dashed-dotted lines correspond to the case as $a \to 0$ of a uniform condensate.}
  \label{fig:free_energy}
\end{figure}

We solve the hydrodynamic boundary value problem using the method of images (Appendix~\ref{app:greensf}). The free energy $E = 2 n_0 \int_{\Omega} |\nabla \chi_{\boldsymbol{\rho}}|^2$ of a vortex (i.e. the additional energy due to the presence of the vortex) at position $\boldsymbol{\rho}$ (radial displacement of $\rho$) in a uniform annular BEC of density $n_0$ occupying the area $\Omega$ is then given by (see Appendix~\ref{app:free_energy})
\begin{equation}
\label{eqn:2en}
\begin{split}
\frac{E}{4\pi n_0} & = \ln{\left(\frac{b}{\xi}\right)} + g(\rho,\rho),
\end{split}
\end{equation}
where $\xi$ is the healing length, and 
\begin{equation}
\label{eqn:2en122}
\begin{split}
g(\eta_1,\eta_2) &= \frac{\ln{\left(\frac{\rho_1}{a}\right)}\ln{\left(\frac{b}{\rho_2}\right)}}{\ln{\left(\frac{b}{a}\right)}} + \ln{\left(\frac{|a^2 - \eta_1\bar{\eta}_2||b^2-\eta_1\bar{\eta}_2|}{b|b^2 \eta_1 - a^2 \eta_2 |}\right)} \\
& + 2\sum_{n=1}^\infty {\frac{a^{2n}(b^{2n}-\rho_2^{2n})(\rho_1^{2n}-a^{2n})}{2n b^{2n} (\rho_1 \rho_2)^{n} (b^{2n} - a^{2n})} \cos{\left( n \phi_{12} \right)}},
\end{split}
\end{equation}
where the bar denotes complex conjugation. All the geometry-related information is in the function $g$. We have defined $\eta_1 = \rho_1 \rme^{\rmi \phi_1}$, $\eta_2 = \rho_2 \rme^{\rmi \phi_2}$, $\lbrace \rho_1, \rho_2 \rbrace \in [a,b]; \lbrace \phi_1,  \phi_2 \rbrace \in [0,2\pi[$, and $\phi_{12} \equiv \phi_1-\phi_2$. Equation~\eqref{eqn:2en} is given to logarithmic accuracy, i.e. we assume $\ln{(b/\xi) \gg 1}$, and that terms of order $\mathcal{O}(\xi^2)$ can be neglected. In the limit as $a \to 0$, Eq.~\eqref{eqn:2en} simplifies to the well-known result~\cite{Fetter98,PhysRev.161.189}
\begin{equation}
\label{eqn:3}
\lim_{a\to 0} E = 4\pi n_0 \left[ \ln{\left(\frac{b}{\xi}\right)} + \ln{\left(1 - \frac{\rho^2}{b^2}\right)}\right]
\end{equation}
for a uniform circular condensate of radius $b$ with an off-axis vortex at $r = \rho$.

For a uniform annular condensate with $N_v$ vortices at $\lbrace \boldsymbol{\rho}_j \rbrace$ of charges $\lbrace \kappa_j \rbrace$ respectively, we obtain
\begin{equation}
\label{eqn:2a2}
\begin{split}
\frac{E}{ 4\pi n_0} &=   N_v \ln{\left(\frac{b}{\xi}\right)} +\sum_{k = 1}^{N_v} g(\rho_k,\rho_k) + \sum_{k < l}^{N_v} \kappa_k \kappa_l \left\lbrace \right. \\
&\,  \Theta(\rho_k \leq \rho_l) \left[\ln{\left(\frac{b}{|\boldsymbol{\rho}_k - \boldsymbol{\rho}_l|} \right)} + g(\eta_k,\eta_l)\right] +\\
&\left. \Theta(\rho_k > \rho_l)\left[\ln{\left(\frac{b}{|\boldsymbol{\rho}_k - \boldsymbol{\rho}_l|} \right)} + g(\eta_l,\eta_k)\right] \right\rbrace,
\end{split}
\end{equation}
where $\Theta$ is the Heaviside step function, and $\phi_{kl} \equiv \phi_k - \phi_l$ is the angle between the $k$th and $l$th vortex. Equation~\eqref{eqn:2a2} agrees with the result obtained by Fetter~\cite{PhysRev.153.285} in terms of the Jacobi theta functions; here, the physical interpretation is transparent and follows directly from the separately given contributions of the annular geometry (functions $g$) and the non-geometry related vortex terms.

\begin{figure}
  \centering
  \includegraphics[width=0.23\textwidth]{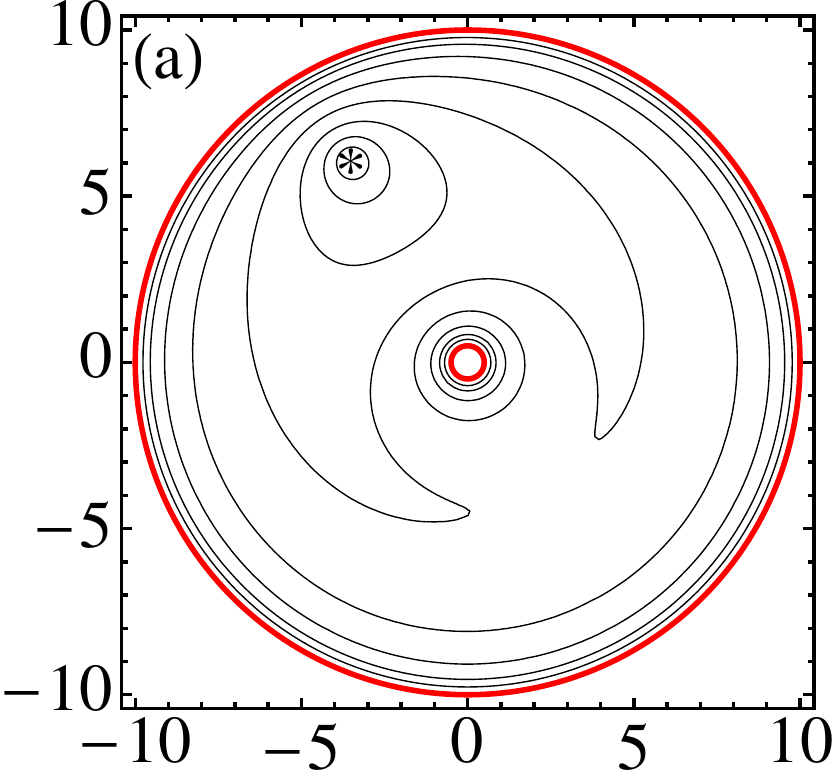}
  \includegraphics[width=0.23\textwidth]{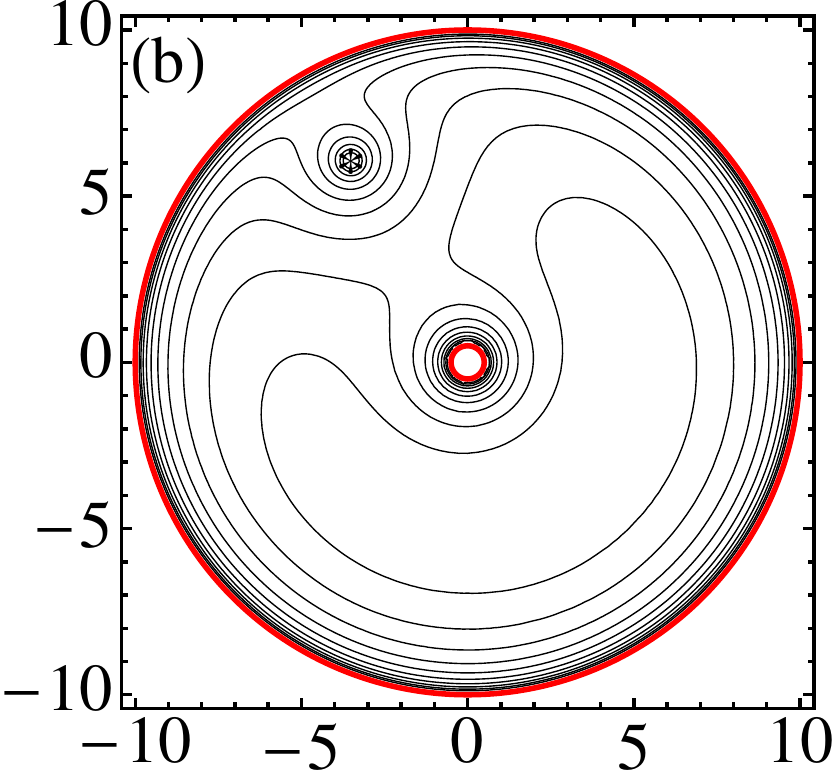}
  \includegraphics[width=0.23\textwidth]{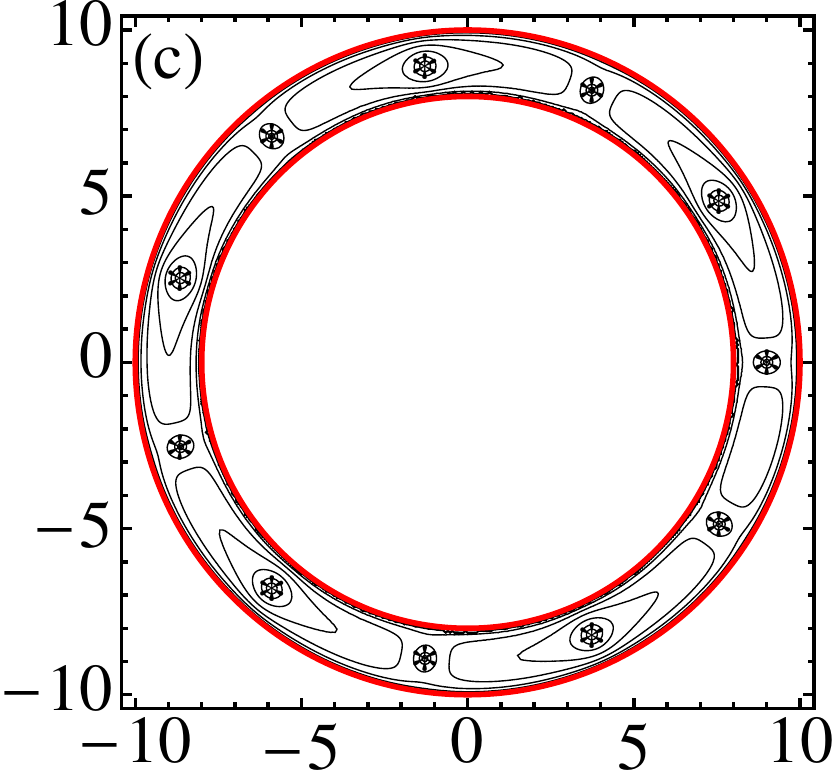}
  \includegraphics[width=0.23\textwidth]{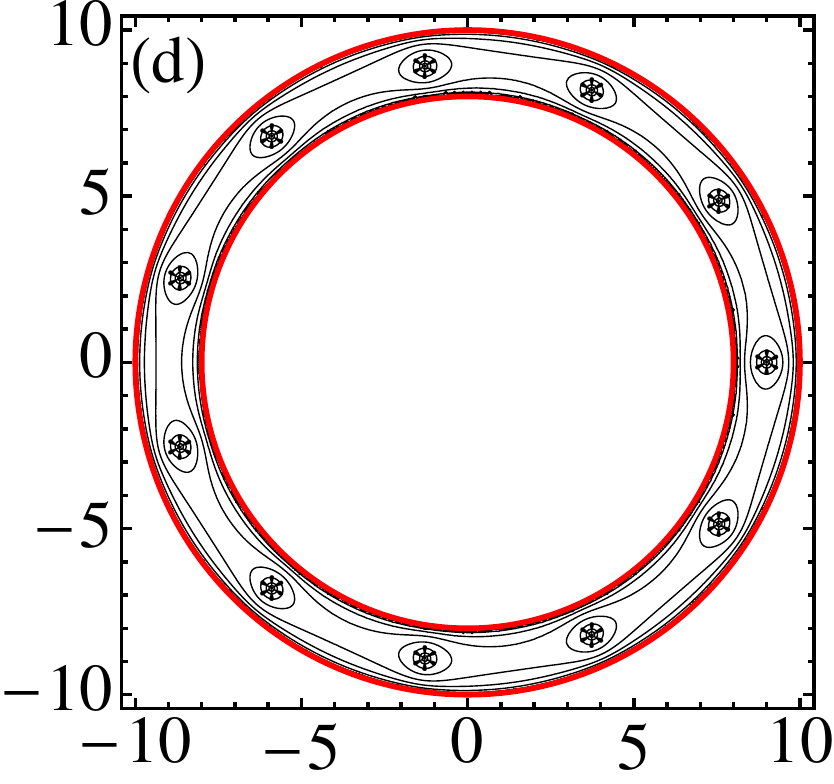}
  \put(-127,-10){x ($\mathrm{a}_{\mathrm{osc}}$)}
  \put(-245,80){\rotatebox{90}{y ($\mathrm{a}_{\mathrm{osc}}$)}}
  \caption{Trajectories of a single vortex. (a) Free energy contours of two equal vortices such that one vortex is held fixed at $r = 7$ and $\phi = 2\pi/3$ (*) with $a = 0.5$ and $b = 10$ [see Eq.~\eqref{eqn:2a2}]. (b) Same as (a) but with opposite vortices. (c) 12 alternating vortices with 11 held fixed in a necklace of radius $9$ ($a = 8$ and $b = 10$). (d) Same as (c) but all the vortices are equal. The red thick curves show the condensate boundary.}
\label{fig:free_energy_cont}
\end{figure}

In the Thomas-Fermi approximation, valid for large strongly-repulsive condensates ($b/\xi \gg 1$), we assume that the spatial variation of the density is so small that the kinetic energy comes only from the velocity, i.e. gradient of the phase. In particular, in this work we assume that $\nabla n_0 \approx \textbf{0}$, and the density at the vortex can be replaced by the vortex-free state, as corrections are of order $\mathcal{O}(\xi^2/b^2)$, and the vortex does not alter the condensate density significantly. Denoting a disk of radius $\xi$ centered at the vortex at $\boldsymbol{\rho}$ by $B_{\boldsymbol{\rho}}^\xi$, the free energy depends only on surface terms $\partial B_{\boldsymbol{\rho}}^\xi$ if we ignore the volume term arising from $\nabla n_0 \neq \textbf{0}$. Therefore, we need only multiply the free energy by the Thomas-Fermi density $n_{\mathrm{TF}}(\boldsymbol{\rho})$ at the vortex:
\begin{equation}
\label{eqn:4p}
\begin{split}
E^{\mathrm{TF}} &= \frac{1}{2} \int_{\Omega \setminus B_{\boldsymbol{\rho}}^\xi} n_{\mathrm{TF}}|v|^2 = \left[1 - \left( \frac{\rho-r_{\mathrm{trap}}}{b}\right)^2\right]E.
\end{split}
\end{equation}

We have plotted the free energy of a single vortex [see Eq.~\eqref{eqn:2en}] in Fig.~\ref{fig:free_energy}(a). The assumptions underlying an incompressible superfluid break down when the vortex is within a few healing lengths from the borders. 

\subsection{\label{sec:2b}Vortex precession in an annulus}
As the Gross-Pitaevskii equation is Hamiltonian in structure, at $T = 0$ in absence of dissipation, a vortex will follow equipotential contours of the free energy, which in general depend on density inhomogeneities. We have plotted contours of the $N_v$-vortex free energy in Fig.~\ref{fig:free_energy_cont}. They are the trajectories of a single (positive) vortex if the other vortices are held fixed. In general, the $N_v$ vortices will follow simultaneous trajectories that leave $E$ time-invariant.

As the free energy of a single vortex depends only on the radial coordinate, the cylindrical symmetry means the vortex will precess around the trap centre, and the rate can be calculated using the Magnus force,
\begin{equation}
\label{eqn:Mag}
n m (\boldsymbol{\kappa} \times \dot{\boldsymbol{\rho}}) = \nabla_{\boldsymbol{\rho}} E,
\end{equation}
where $n$ is the number density and $\boldsymbol{\kappa} = (h \kappa / m)\hat{\textbf{z}}$ is the circulation vector of a vortex of charge $\kappa \in \field{Z}$. Using Eqs.~\eqref{eqn:2en} and~\eqref{eqn:4p}, in a uniform condensate in the annulus $\Omega$, an off-axis vortex at $r = \rho$ will precess at the rate
\begin{equation}
\label{eqn:uniform_annulus}
\begin{split}
\omega =& - 2 \frac{\ln{\left(\frac{b}{\rho}\right)}-\ln{\left(\frac{\rho}{a}\right)}}{\rho^2 \ln{\left(\frac{b}{a}\right)}} + \frac{2}{\rho^2} - \frac{4}{\rho^2-a^2} -\frac{4}{\rho^2 - b^2}  \\
&  -\frac{4}{\rho^2} \sum_{n=1}^\infty \frac{\left(\frac{a}{\rho}\right)^{2n}-\left(\frac{\rho}{b}\right)^{2n}}{\left(\frac{b}{a}\right)^{2n} - 1} ,
\end{split}
\end{equation}
while in an annular Thomas-Fermi condensate $n_{\mathrm{TF}} = n_0\left[1 - \left( \frac{r-r_{\mathrm{trap}}}{b}\right)^2\right]$, the precession rate is
\begin{equation}
\label{eqn:TF_annulus}
\omega_{\mathrm{TF}} = \omega + \frac{4}{b^2} \frac{1- \frac{r_{\mathrm{trap}}}{\rho}}{1 - \left( \frac{\rho-r_{\mathrm{trap}}}{b}\right)^2} \left[ \ln{\left(\frac{b}{\xi}\right)} + g(\rho,\rho)\right].
\end{equation}

We show plots of Eqs.~\eqref{eqn:uniform_annulus} and~\eqref{eqn:TF_annulus} in Fig.~\ref{fig:free_energy}(b). It can be seen in Fig.~\ref{fig:free_energy}(b) that the annular boundary can formally stabilise a vortex (i.e. $\omega = 0$), which is not the case in cylindrical geometry. The precession frequency changes sign at this quasi-stable radius. Similar results have been obtained in Ref.~\cite{PhysRevA.66.033602} using numerical integration of the Gross-Pitaevskii equation and the method of images, and an approximative analytical solution of the Poisson equation, where the annulus is conformally mapped onto a straight strip. In comparison, our solution for the stream function is an exact Green's function of the Dirichlet problem in the annulus.

Instead of the Magnus force approach of Eq.~\eqref{eqn:Mag}, the precession around the ring could have been obtained by numerically solving equations of motion arising from the method of images, updating the most current image configuration at every time step~\cite{PhysRevA.66.033602}. At first sight it may seem strange that the Magnus force and the method of images give equivalent results, as any equivalence might be expected to be ruled out because the method of images concerns the boundaries of the condensate, while the Magnus force is a property of the vortex itself~\cite{PhysRevLett.70.2158} present even in an infinite homogeneous system. However, the boundaries affect the free energy of a vortex (the functions $g$), and when there is no dissipation, the vortex must follow contours of the free energy. Therefore, when the vortex moves, the transverse Magnus force must become exactly the force arising from the gradient of the free energy, and thus be affected by the boundaries. The method of images, on the other hand, works because by the Uniqueness Theorem any solution to the Poisson equation with given boundary conditions [see Eq.~\eqref{eqn:Poisson_stream}] is unique. For an incompressible superfluid in a flat-bottom trap, therefore, the free energy contours can thus be obtained either with the method of images, or by considering the Magnus force.

\subsection{\label{sec:2c}Phonon radiation by a precessing vortex} 
\begin{figure}
  \centering
  \includegraphics[width=0.45\textwidth]{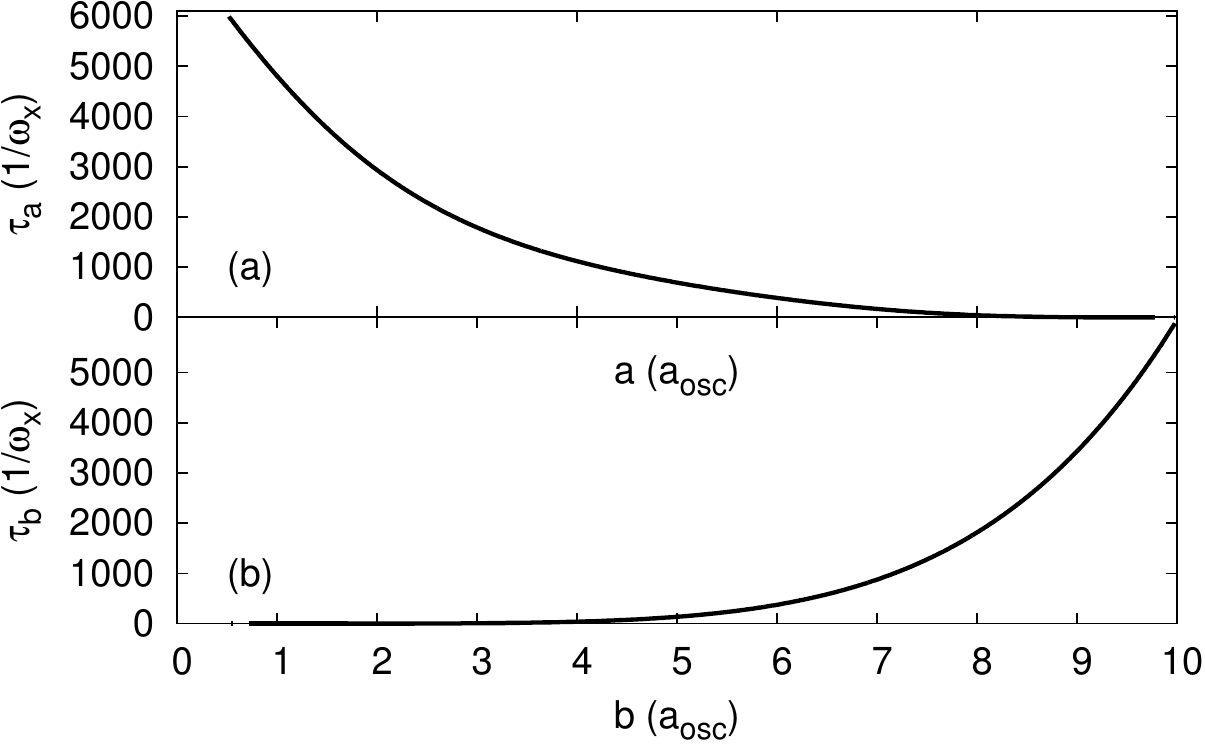}
  \caption{Vortex lifetime. (a) Lower bound for the lifetime of a vortex with $q = \pm 1$ drifting towards the inner edge $r = a$. Here $b = 10.0$. (b) Same as (a) but for a vortex drifting towards the outer edge $r = b$. Here $a = 0.5$. In both figures $\xi = 0.01$, and we have scaled $c = \sqrt{\pi}/2$.}
  \label{fig:lifetime}
\end{figure}

In Ref.~\cite{PhysRevA.61.063612}, the power $P$ radiated by a vortex executing circular motion in an infinite uniform system was obtained to be
\begin{equation}
\label{eqn:power_radiated}
P = \frac{\pi q_v^2 \omega^3 \rho^2}{4c^2},
\end{equation}
where $q_v = -\hbar \sqrt{2\pi n_0/m}$, and the speed of sound $c = \sqrt{g n_0 / m}$. Setting $P = \mathrm{d}E/\mathrm{d}t = (\partial E/\partial \rho) \mathrm{d}\rho/\mathrm{d}t$ and using Eqs.~\eqref{eqn:2en} and~\eqref{eqn:uniform_annulus}, we may obtain a lower bound for the vortex lifetimes $\tau_{\lbrace a,b\rbrace }$ in the annulus:
\begin{equation}
\label{eqn:lifetime}
\tau_{\lbrace a,b\rbrace } = \left| \int_{r_{\mathrm{\omega}}\pm 10\xi}^{\lbrace a,b\rbrace \mp \xi} \frac{\partial E}{\partial \rho} \frac{\mathrm{d}\rho}{P(\rho)} \right| = \frac{4c^2}{\pi q_v^2} \left|  \int_{r_{\mathrm{\omega}}\pm 10\xi}^{\lbrace a,b\rbrace \mp \xi} \frac{\partial E}{\partial \rho} \frac{\mathrm{d}\rho}{\omega^3 \rho^2} \right|,
\end{equation}
where $r_{\mathrm{\omega}}$ is the radius corresponding to $\omega = 0$. Numerically evaluating the integral (Fig.~\ref{fig:lifetime}) shows that while the vortex lifetimes remain small for very narrow condensates, the vortex quickly becomes long-lived as the thickness of the annulus is increased. Decreasing the healing length also makes the vortex lifetime longer.

\section{Conclusions} 
We have studied vortex dynamics in an annular BEC, obtaining expressions for the free energy, precession rates, and vortex lifetimes. The results form an exact solution to the incompressible hydrodynamic boundary value problem, and provide closed-form expressions for key observable features of vortex dynamics in an annulus. The incompressible hydrodynamics presented here forms the starting point for more generalised discussions of vortex dynamics in spin-orbit coupled and compressible superfluids~\cite{LAT_JB_15} in non-simply connected geometries, which will be discussed later.  

We acknowledge the support of the Jenny and Antti Wihuri Foundation. We would like to thank A. Fetter for inspiring discussions. LAT would like to thank the Turku Centre for Quantum Physics for hospitality.

\appendix

\section{Green's function in an annulus}
\label{app:greensf}
In this Appendix, we derive the Green's function in an annulus $\Omega = \{a < r < b; 0 \leq \varphi < 2\pi\}$ with Dirichlet boundary conditions using the method of images. By definition, the Green's function satisfies the Poisson equation
\begin{equation}
\nabla^2 G_{\boldsymbol{\rho}}(\textbf{r}) = 2\pi \delta(\textbf{r} - \boldsymbol{\rho}),
\end{equation}
where $\delta$ is a two-dimensional delta function, and $\left\lbrace \textbf{r}, \boldsymbol{\rho}\right\rbrace \in \Omega$. In general, the Green's function can be decomposed into a radially symmetric singular and a regular part, which we label by $F_{\boldsymbol{\rho}}(\mathbf{r})$ and $H_{\boldsymbol{\rho}}(\mathbf{r})$ respectively:
\begin{equation}
\label{eqn:G_decomp}
G_{\boldsymbol{\rho}}(\textbf{r}) = F_{\boldsymbol{\rho}}(\mathbf{r}) + H_{\boldsymbol{\rho}}(\mathbf{r}).
\end{equation}
The singular part $F_{\boldsymbol{\rho}}(\mathbf{r})$ is the fundamental solution corresponding to system response at a general point $\mathbf{r}$ to a point source at some arbitrary point $\boldsymbol{\rho}$, and can be written down as
\begin{equation}
F_{\boldsymbol{\rho}}(\mathbf{r}) = -\frac{1}{2}\ln{\left(\left|\mathbf{r}-\boldsymbol{\rho} \right|^2 \right)}.
\end{equation}
The regular part $H_{\boldsymbol{\rho}}(\mathbf{r})$ is harmonic in $\Omega$ satisfying Laplace's equation there. The choice of this analytic harmonic function is asserted by the boundary conditions. Installing $N$ images at $\lbrace \boldsymbol{\tilde{\rho}}_j \rbrace$, all outside of $\Omega$, we can write
\begin{equation}
H_{\lbrace \boldsymbol{\tilde{\rho}}_j \rbrace}(\mathbf{r}) = \sum_{j=1}^N (\pm)_j \frac{1}{2}\ln{\left(\left|\mathbf{r}-\boldsymbol{\tilde{\rho}}_j \right|^2 \right)},
\end{equation}
where the sign depends on the winding number of the particular image vortex.

For the annulus, we need two sets of images that both have an infinite number of image vortices~\cite{PhysRevA.64.063602,PhysRevA.66.033602,Chen2009678}. Importantly, the limit of infinitely many images must be taken with care at the origin when constructing the Green's function below. The image sets are constructed by considering the outer and inner surfaces of the annulus as separate circular boundaries, whose effect on each other must be compensated by a limit of having two infinite series of images. Assuming a vortex with circulation $\kappa$ at $\boldsymbol{\rho}$, the outer surface induces an image at $\boldsymbol{\tilde{\rho}}_1 = (b^2 / \rho)\hat{\boldsymbol{\rho}}$ of circulation $-\kappa$. While temporarily fixing the boundary condition at the outer surface, the inner surface must be compensated with another image at $\boldsymbol{\tilde{\rho}}_2 = (a^2 / b^2)\boldsymbol{\rho}$ of circulation $\kappa$, as well as an image vortex at the origin of circulation $-\kappa$~\cite{Saffman97}, which then affect the outer surface again, and so on. A similar chain of images is needed starting from the inner surface.

Considering a stream function $\chi_{\boldsymbol{\rho}}(\textbf{r})$ such that the superfluid velocity $\textbf{v}$ is given by
\begin{equation}
\label{eqn:sf_vel}
\textbf{v}(\textbf{r}) = \frac{\hbar}{m}\hat{\textbf{z}} \times \nabla \chi_{\boldsymbol{\rho}}(\textbf{r}),
\end{equation}
it follows by definition of the vorticity $\nabla \times \textbf{v}$ of irrotational flow that the stream function satisfies the Poisson equation. To take into account the physical no-flow requirement $n \textbf{v}\cdot \hat{\textbf{n}} = 0$ and the geometry of the system, where $n$ is the superfluid density and $\hat{\textbf{n}}$ the unit outward normal, we require Dirichlet boundary conditions. In the annulus $\Omega$ with $N_v$ vortices at $\lbrace \boldsymbol{\rho}_j \rbrace\, (j = 1,2,\ldots N_v)$, we have:
\begin{equation}
\begin{cases}
\nabla^2 \chi_{\lbrace \boldsymbol{\rho}_j \rbrace}(\textbf{r}) = 2\pi \sum_{j=1}^{N_v} \delta(\textbf{r}-\boldsymbol{\rho}_j), \\
\chi_{\lbrace \boldsymbol{\rho}_j \rbrace}(a) = \chi_{\lbrace \boldsymbol{\rho}_j \rbrace}(b) = 0.
\end{cases}
\end{equation}

\begin{figure}
  \centering
  \includegraphics[width=0.15\textwidth]{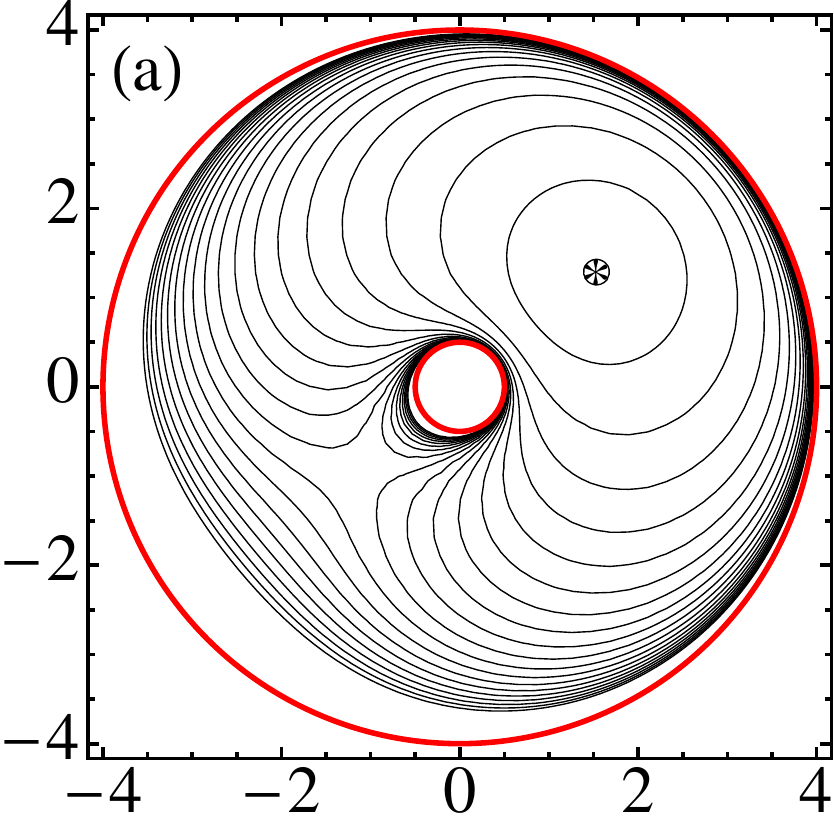}
  \includegraphics[width=0.15\textwidth]{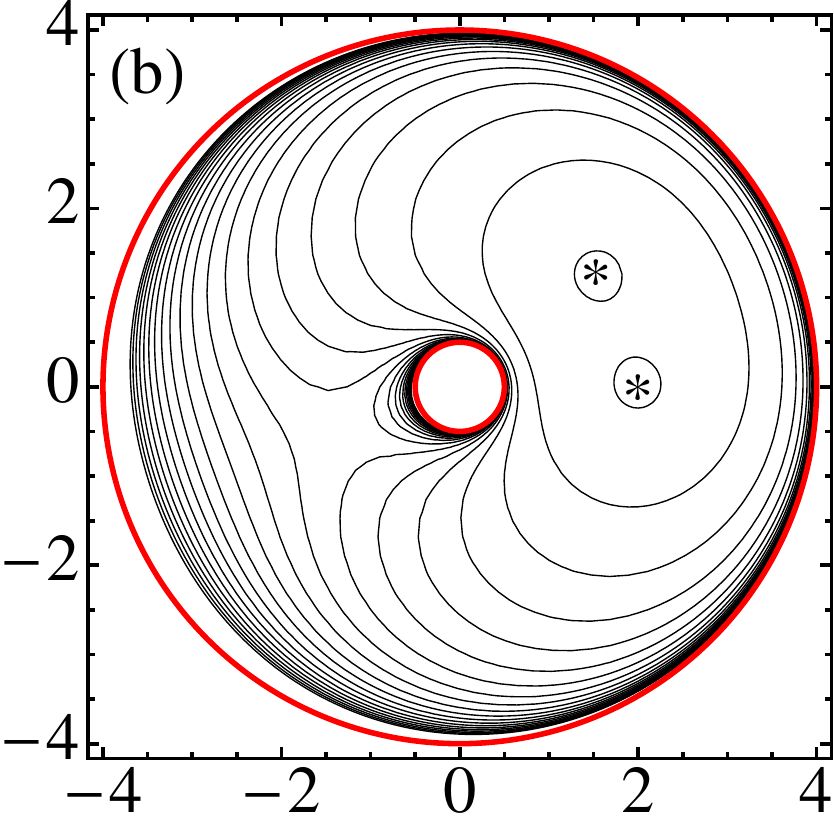}
  \includegraphics[width=0.15\textwidth]{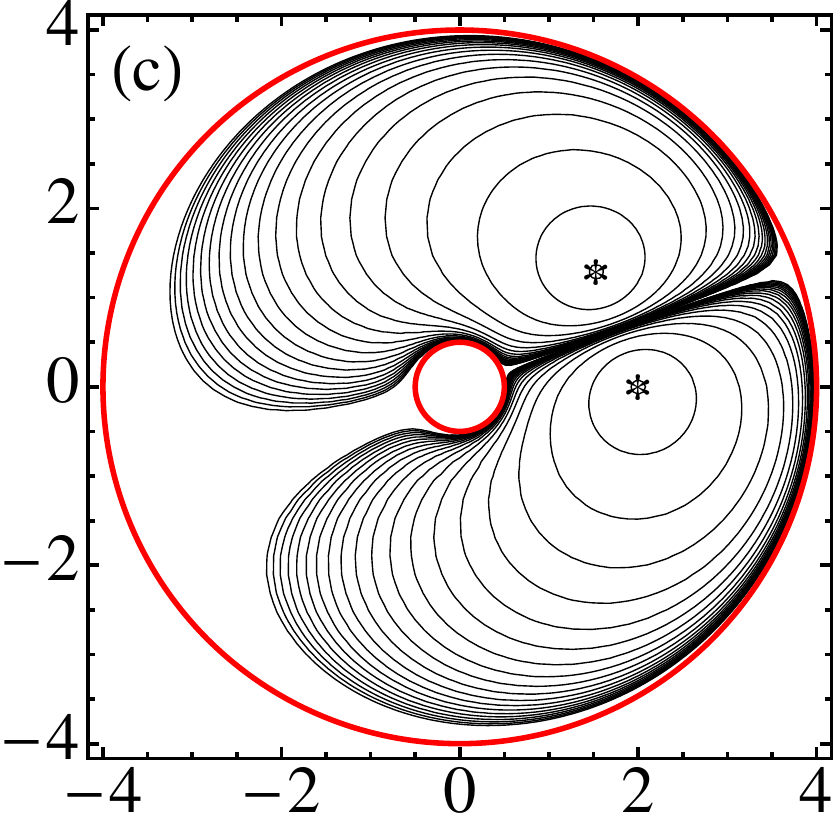}
  \caption{Annular BEC with $a = 0.5$, and $b = 4.0$. (a) The stream function $\chi$ [i.e. constant-velocity flow lines, see Eq.~\eqref{eqn:sf}] of a single vortex (*) at $\rho = 2.0$. (b) Two (positive) vortices at $\rho = 2.0$ separated by an angle of $2\pi/9$. (c) Same as (b) but with two opposite vortices. The red circles show the condensate boundaries.}
  \label{fig:Streamfn}
\end{figure}

Inserting the images as per the Dirichlet boundary conditions, we obtain
\begin{equation}
\label{eqn:chi1}
\begin{split}
\chi_{\boldsymbol{\rho}}(\textbf{r}) & = \frac{1}{2}\ln{\left(\frac{1}{\left|\mathbf{r}-\boldsymbol{\rho} \right|^2 }\right)} + \sum_{j=1}^\infty (\pm)_j \frac{1}{2}\ln{\left(\left|\mathbf{r}-\boldsymbol{\tilde{\rho}}_j \right|^2 \right)} \\
& \stackrel{N\to\infty}{=} \frac{1}{2}\ln{\left(\frac{1}{\left|\mathbf{r}-\boldsymbol{\rho} \right|^2 }\right)} \\
&- 2 N \ln{\left(\frac{\rho}{a}\right)} - \frac{\ln{\left(\frac{\rho}{a}\right)} \ln{\left(b\right)} +\ln{\left(\frac{b}{\rho}\right)} \ln{\left(r\right)}}{\ln{\left(\frac{b}{a}\right)}} \\
& - \frac{1}{2}\sum_{j=1}^N \left\lbrace -\ln{\left[\left|\mathbf{r}-\left(\frac{a}{b} \right)^{2j}\frac{b^2}{\rho} \boldsymbol{\hat{\rho}} \right|^2 \right]} - (a \leftrightarrow b) \right. \\
 & \qquad \left. +\ln{\left[\left|\mathbf{r}-\left(\frac{b}{a} \right)^{2j}\rho \boldsymbol{\hat{\rho}} \right|^2 \right]} + (a \leftrightarrow b) \right\rbrace.
\end{split}
\end{equation}
Equation~\eqref{eqn:chi1} for $\chi_{\boldsymbol{\rho}}(\textbf{r})$ can be regrouped to read
\begin{equation}
\label{eqn:sf}
\begin{split}
& \chi_{\boldsymbol{\rho}}(\textbf{r})  = \Theta(r \geq \rho) \left\lbrace \frac{\ln{\left(\frac{\rho}{a}\right)}\ln{\left(\frac{b}{r}\right)}}{\ln{\left(\frac{b}{a}\right)}} + \ln{\left[\frac{|a^2 - z\bar{\eta}||b^2-z\bar{\eta}|}{|z-\eta | |b^2 \eta - a^2 z |}\right]} \right. \\
 &   \qquad \left. + 2\sum_{n=1}^\infty {\frac{a^{2n}(b^{2n}-r^{2n})(\rho^{2n}-a^{2n})}{2n b^{2n} (\rho r)^{n} (b^{2n} - a^{2n})} \cos{\left[n(\varphi-\phi) \right]}}\right\rbrace \\
& + \Theta(r < \rho)\left\lbrace \frac{\ln{\left(\frac{r}{a}\right)}\ln{\left(\frac{b}{\rho}\right)}}{\ln{\left(\frac{b}{a}\right)}} + \ln{\left[\frac{|a^2 - z\bar{\eta}||b^2-z\bar{\eta}|}{|z-\eta | |b^2 z - a^2 \eta |}\right]} \right. \\
 &  \qquad \left. + 2\sum_{n=1}^\infty {\frac{a^{2n}(b^{2n}-\rho^{2n})(r^{2n}-a^{2n})}{2n b^{2n} (\rho r)^{n} (b^{2n} - a^{2n})}\cos{\left[n(\varphi-\phi) \right]}}\right\rbrace,
\end{split}
\end{equation}
where $z = r \rme^{\rmi \varphi}$, $\eta = \rho \rme^{\rmi \phi}$ is the position of the vortex, and $\Theta$ is the Heaviside step function. We show a few example plots of Eq.~\eqref{eqn:sf} in Fig.~\ref{fig:Streamfn}.

In the limit as $a \to 0$ of a circular trap, the stream function in Eq.~\eqref{eqn:sf} is given by
\begin{equation}
\label{eqn:sfa}
\lim_{a\to 0}\chi_{\boldsymbol{\rho}}(\textbf{r})  = \ln{\left(\frac{|b^2-r \rho \rme^{\rmi (\varphi - \phi)}|}{b|r \rme^{\rmi (\varphi - \phi)}-\rho|}\right)},
\end{equation}
which can be decomposed into the singular and regular parts $\ln{\left(b/|r \rme^{\rmi (\varphi - \phi)}-\rho|\right)}$ and $\ln{\left(\left|1-r \rho \rme^{\rmi (\varphi - \phi)}/b^2\right|\right)}$, respectively.

\section{Evaluation of the free energy}
\label{app:free_energy}
The free energy of a vortex at position $\boldsymbol{\rho}$ in a uniform annular BEC of density $n_0$ occupying the area $\Omega = \{a < r < b; 0 \leq \varphi < 2\pi\}$ is:
\begin{equation}
\label{eqn:1}
E = \frac{1}{2}n_0 \int_{\Omega} |v|^2 = \infty.
\end{equation}
The diverging part arises from the vortex core, and does not contain physically interesting information. Let us therefore regulate the kinetic energy by excising $B_{\boldsymbol{\rho}}^\xi$, a small disk of radius $\xi$ centered at the vortex, where $\xi$ is the healing length. Using Eqs.~\eqref{eqn:sf_vel} and~\eqref{eqn:G_decomp}, $\nabla \cdot (n \chi \nabla \chi) = (\nabla n)\cdot (\chi \nabla\chi)+ n|\nabla \chi|^2 + n\chi\nabla^2\chi$, $\nabla^2\chi = 2\pi \delta(\textbf{r}-\boldsymbol{\rho})$, and the Divergence Theorem noting that the singularity is not inside the integration domain, the finite kinetic energy thus obtained is
\begin{equation}
\label{eqn:2a}
\begin{split}
E &= \frac{1}{2}n_0 \int_{\Omega \setminus B_{\boldsymbol{\rho}}^\xi} |v|^2 = 2n_0 \int_{\Omega \setminus B_{\boldsymbol{\rho}}^\xi} |\nabla \chi_{\boldsymbol{\rho}}(\textbf{r}) |^2 \\
& = -2n_0 \int_{\partial B_{\boldsymbol{\rho}}^\xi} (F_{\boldsymbol{\rho}}+H_{\boldsymbol{\rho}}) \partial_n (F_{\boldsymbol{\rho}}+H_{\boldsymbol{\rho}}),
\end{split}
\end{equation}
where $\partial_n$ denotes the outward unit normal derivative on $\partial B_{\boldsymbol{\rho}}^\xi$. Using Eq.~\eqref{eqn:sf}, the individual terms are readily evaluated to yield
\begin{equation}
\label{eqn:2}
\begin{split}
E & = 4\pi n_0 \left\lbrace \ln{\left(\frac{b}{\xi}\right)} + \frac{\ln{\left(\frac{\rho}{a}\right)}\ln{\left(\frac{b}{\rho}\right)}}{\ln{\left(\frac{b}{a}\right)}} \right. \\
& \qquad \qquad + \ln{\left[\frac{(\rho^2-a^2)(b^2-\rho^2)}{b\rho (b^2 - a^2)}\right]} \\
&  \qquad \qquad \left. + 2\sum_{n=1}^\infty {\frac{a^{2n}(b^{2n}-\rho^{2n})(\rho^{2n}-a^{2n})}{2n b^{2n} \rho^{2n} (b^{2n} - a^{2n})}}\right\rbrace.
\end{split}
\end{equation}
This result is given to logarithmic accuracy, i.e. we assume $\ln{(b/\xi) \gg 1}$, and that terms of order $\mathcal{O}(\xi^2)$ can be neglected.

Having multiple vortices in the condensate only adds non-singular interaction terms of the form $\int_{\Omega \setminus \lbrace B_{\boldsymbol{\rho}_i}^\xi \rbrace} \nabla \chi_{\boldsymbol{\rho}_l} \cdot \nabla \chi_{\boldsymbol{\rho}_k}$, which can be evaluated using similar methods as above. For a uniform annular condensate with $N_v$ vortices at $\boldsymbol{\rho} = \lbrace \boldsymbol{\rho}_j \rbrace$ of charges $\lbrace \kappa_j \rbrace$ respectively, we obtain
\begin{widetext}
\begin{equation}
\label{eqn:2a22}
\begin{split}
&\frac{E}{ 4\pi n_0} = \,  N_v \ln{\left(\frac{b}{\xi}\right)} +\sum_{k = 1}^{N_v} \left\lbrace \frac{\ln{\left(\frac{\rho_k}{a}\right)}\ln{\left(\frac{b}{\rho_k}\right)}}{\ln{\left(\frac{b}{a}\right)}} + \ln{\left[\frac{(\rho_k^2-a^2)(b^2-\rho_k^2)}{b\rho_k (b^2 - a^2)}\right]} + 2\sum_{n=1}^\infty {\frac{a^{2n}(b^{2n}-\rho_k^{2n})(\rho_k^{2n}-a^{2n})}{2n b^{2n} \rho_k^{2n} (b^{2n} - a^{2n})}}\right\rbrace + \sum_{k < l}^{N_v} \kappa_k \kappa_l \left\lbrace \right. \\
&\,  \Theta(\rho_k \leq \rho_l) \left[\frac{\ln{\left(\frac{\rho_k}{a}\right)}\ln{\left(\frac{b}{\rho_l}\right)}}{\ln{\left(\frac{b}{a}\right)}} + \ln{\left( \frac{|a^2 - \rho_k \rho_l \rme^{\rmi\phi_{kl}}||b^2 - \rho_k \rho_l  \rme^{\rmi\phi_{kl}}|}{|\boldsymbol{\rho}_k - \boldsymbol{\rho}_l||b^2 \rho_k \rme^{\rmi\phi_{kl}} - a^2 \rho_l|} \right)}  + 2\sum_{n = 1}^\infty{\frac{a^{2n}(b^{2n}-\rho_l^{2n})(\rho_k^{2n}-a^{2n})}{2n b^{2n} (\rho_l \rho_k)^{n} (b^{2n} - a^{2n})} \cos{(n \phi_{kl})}} \right] \\
&\left. + \Theta(\rho_k > \rho_l)\left[ \frac{\ln{\left(\frac{\rho_l}{a}\right)}\ln{\left(\frac{b}{\rho_k}\right)}}{\ln{\left(\frac{b}{a}\right)}} + \ln{\left( \frac{|a^2 - \rho_k \rho_l \rme^{\rmi\phi_{kl}}||b^2 - \rho_k \rho_l  \rme^{\rmi\phi_{kl}}|}{|\boldsymbol{\rho}_k - \boldsymbol{\rho}_l||b^2 \rho_l - a^2 \rho_k \rme^{\rmi\phi_{kl}}|} \right)}  + 2\sum_{n = 1}^\infty{\frac{a^{2n}(b^{2n}-\rho_k^{2n})(\rho_l^{2n}-a^{2n})}{2n b^{2n} (\rho_l \rho_k)^{n} (b^{2n} - a^{2n})} \cos{(n \phi_{kl})}} \right]    \right\rbrace,
\end{split}
\end{equation}
\end{widetext}
where $\phi_{kl} \equiv \phi_k - \phi_l$ is the angle between the $k$th and $l$th vortex. 

When $a \to 0$, Eq.~\eqref{eqn:2a22} simplifies to
\begin{equation}
\label{eqn:2b}
\begin{split}
\frac{E}{4\pi n_0} &\stackrel{a \to 0}{=} \,   N_v \ln{\left(\frac{b}{\xi}\right)} + \sum_{k = 1}^{N_v} \ln{\left(1-\frac{\rho_k^2}{b^2}\right)} \\
&\, + \frac{1}{2} \sum_{k < l}^{N_v} \kappa_k \kappa_l \ln{\left[ \frac{ b^2 - 2\rho_k \rho_l\cos{(\phi_{kl})} + \left( \frac{\rho_k \rho_l}{b}\right)^2 }{\rho_k^2 - 2\rho_k \rho_l \cos{(\phi_{kl})} + \rho_l^2} \right]}.
\end{split}
\end{equation}

\bibliographystyle{apsrev4-1}
\bibliography{references}

\end{document}